\documentclass[fleqn,10pt]{SelfArx} 

\usepackage[english]{babel} 

\usepackage{lipsum} 
\usepackage{afterpage}

 \definecolor{color1}{RGB}{0,0,110} 
\definecolor{color2}{RGB}{0,20,80} 

\usepackage{hyperref} 
\hypersetup{hidelinks,colorlinks,breaklinks=true,urlcolor=color2,citecolor=color1,linkcolor=color1,bookmarksopen=false,pdftitle={Title},pdfauthor={Author}}

\JournalInfo{Uploaded to arXiv.org} 
\Archive{} 

\PaperTitle{Multiscale modeling of a rectifying bipolar nanopore: explicit-water versus implicit-water simulations} 

\Authors{Zolt\'an Hat\'o\textsuperscript{1}, M\'onika Valisk\'o\textsuperscript{1}, Tam\'as Krist\'of\textsuperscript{1},  Dirk Gillespie\textsuperscript{2},  Dezs\H{o} Boda\textsuperscript{1,3}*} 
\affiliation{\textsuperscript{1}\textit{Department of Physical Chemistry, University of Pannonia,  P. O. Box 158, H-8201 Veszpr\'em, Hungary}} 
\affiliation{\textsuperscript{2}\textit{Department of Physiology and Biophysics, Rush University Medical Center, Chicago, IL 60612}} 
\affiliation{\textsuperscript{3}\textit{Institute of Advanced Studies K\"oszeg (iASK), Chernel u. 14, H-9730 K\"oszeg, Hungary}} 
\affiliation{*\textbf{Corresponding author}: dezsoboda@gmail.com} 

\newcommand{\keywordname}{Keywords} 

\Abstract{In a multiscale modeling approach, we present computer simulation results for a rectifying bipolar nanopore on two modeling levels.
In an all-atom model, we use explicit water to simulate ion transport directly with the molecular dynamics technique.
In a reduced model, we use implicit water and apply the Local Equilibrium Monte Carlo method together with the Nernst-Planck transport equation. 
This hybrid method makes the fast calculation of ion transport possible at the price of lost details.
We show that the implicit-water model is an appropriate representation of the explicit-water model when we look at the system at the device (i.e., input vs.\ output) level. 
The two models produce qualitatively similar behavior of the electrical current for different voltages and model parameters.
Looking at details of concentration and potential profiles, we find profound differences between the two models.
These differences, however, do not influence the basic behavior of the model as a device because they do not influence the $z$-dependence of the concentration profiles which are the main determinants of current.
These results then address an old paradox: how do reduced models, whose assumptions should break down in a nanoscale device, predict experimental data?
Our simulations show that reduced models can still capture the overall device physics correctly, even though they get some important aspects of the molecular-scale physics quite wrong; reduced models work because they include the physics that is necessary from the point of view of device function.
Therefore, reduced models can suffice for general device understanding and device design, but more detailed models might be needed for molecular level understanding. }

\begin{document}

\flushbottom 

\maketitle 


\thispagestyle{empty} 


\section*{Introduction}
\label{sec:intro}

Our devices are getting smaller and smaller.
Nanodevices that involve ion transport generally have regions that can be considered macroscopic from the molecular point of view, with characteristic size of $\mu$m scale and larger (e.g., wide part of a conical nanopore, access regions, bulk regions with electrodes).
Continuum theories are often sufficient to model and compute devices built of such regions.
These reduced (or coarse-grained) models make approximations like point-charge ions and implicit water as a trade-off for fast computation time.
As the dimensions of the devices shrink, however, nanoscopic regions appear (e.g., narrow bottlenecks of nanopores), where the assumptions of these reduced models are no longer valid and molecular models and particle simulations like all-atom molecular dynamics (MD) are needed to reveal molecular mechanisms.
To connect with experimental data then requires a multiscale modeling approach, where different models/methods with different resolutions are applied to different regions.
In our approach, multiscaling is applied between models for the whole system in separate simulations rather than between separate regions in a single model and simulation.
The multiscaling part of our procedure is connecting the various modeling levels; molecular insights obtained from atomic models, for example, can be used to design better reduced models.

Unexpectedly, however, reduced models often reproduce---and predict---experimental data, even though they should, in principle, not work because their underlying atomic physics is too approximate; one should not assume that water molecules are absent or that ions are point charges when the pore they go through is only a few times wider than the ``real'' ion or water.  
Yet, the Poisson-Nernst-Planck (PNP) theory can reproduce experimental data for rectifying nanopores \cite{cervera_jcp_2006,kalman_am_2008,cheng_acsnano_2009,hoffmann_l_2013,pietschmann_pccp_2013,perezmitta_pcp_2016}.
Other examples include reduced models using hard sphere ions reproducing and predicting nonlinear phenomena in biological ion channels \cite{2000_nonner_bj_1976,gillespie-bj-95-2658-2008,gillespie-bj-2008,boda-jgp-133-497-2009,dirk-janhavi-mike,malasics-bba-1798-2013-2010,gillespie_bj_2014,boda-jpcc-118-700-2014} and in nanopores \cite{gillespie-bj-95-609-2008,he-jacs-131-5194-2009}.
Moreover, in many cases including nanopores and ion channels, reduced models are the only way to connect with experimental results, as the required low ion concentrations and small applied voltages are inaccessible to all-atom MD simulations.

The goal of this work is to report computational results for a nanodevice, a rectifying nanopore, on two distinct modeling levels that differ in the treatment of water.
In one model, water is a continuum background and the ions are charged, hard spheres that move via drift-diffusion.
In the other model, MD simulations with explicit water molecules are used.
This simple case study allows us to identify the effects of using an implicit water model versus having corpuscular water molecules.

Consistent with the work cited above, we find that both models produce qualitatively the same input-output relations (e.g., current vs.\ voltage, pore surface charge, or ion concentration).
On the other hand, the molecular-scale picture inside the nanopore for the two models is quite different, as one would expect.
The solution to this paradox is that both models produce qualitatively the same axial ion concentration profiles (i.e., cross-section averaged profiles), and because these are the first-order determinants of current, the two models produce the same device characteristics.

Our results suggest that reduced models that reproduce experimental results are, in fact, valid because they capture the overall device physics correctly despite the incomplete description of the molecular level phenomena.
Therefore, they are useful for understanding the device-level physics that produces the input-output relations and for device design.
At that level, the molecular details are of second order importance.
However, to understand the physics at an atomic scale (e.g., ion and water structure inside a confining nanopore), more detailed molecular models must be used.

\section*{Models}
\label{sec:models}

In an all-atom model, water molecules are modeled explicitly and the system is simulated with the MD technique. 
The ionic current, driven by an external electric field, is simulated directly by counting the ions that passed through the pore (more detail is found in the Appendix and in the ESI).

Coarse-grained models reduce the number of degrees of freedom in the Hamiltonian by integrating out some of these degrees of freedom and replacing them with a response function, or simplify model potentials describing certain parts of the system.
In this work, the reduced model describes water as an implicit continuum background, where the effect of water is replaced by various response functions:\\
(1) The ability of water to affect the movement of ions through friction (a dynamic effect) is taken into account by a diffusion coefficient, $D_{i}(\mathbf{r})$, in the Nernst-Planck (NP) transport equation with which we compute the flux:
\begin{equation}
 -kT\mathbf{j}_{i}(\mathbf{r}) = D_{i}(\mathbf{r})c_{i}(\mathbf{r})\nabla \mu_{i}(\mathbf{r}),
 \label{eq:np}
\end{equation} 
where $\mathbf{j}_{i}(\mathbf{r})$ is the particle flux density of ionic species $i$, $c_{i}(\mathbf{r})$ is concentration, $k$ is Boltzmann's constant, $T$ is temperature, and $\mu_{i}(\mathbf{r})$ is the electrochemical potential profile. 
Its gradient is the driving force of the steady-state transport.\\
(2) The ability of water to screen the charges of ions (an energetic effect) is taken into account by a dielectric constant, $\epsilon$, in the denominator of the Coulomb-potential acting between the charged hard spheres with which we model the ions:
\begin{equation}
 u_{ij}(r)
=\left\{
        \begin{array}{ll}
    \infty & \; \mbox{for} \; \;  r<R_{i}+R_{j}\\
        \dfrac{q_{i}q_{j}}{4\pi \epsilon_{0}\epsilon r} & \; \mbox{for} \; \; r \geq R_{i}+R_{j}  ,
        \end{array}
        \right. 
\label{eq:pm}
\end{equation} 
where $q_{i}$ is the charge and $R_{i}$ is the radius of ionic species $i$, $\epsilon_{0}$ is the permittivity of vacuum, and $r$ is the distance between the ions.

We simulate this model with the Local Equilibrium Monte Carlo (LEMC) technique \cite{boda-jctc-8-824-2012,boda-jml-189-100-2014,boda-arcc-2014}, which is practically a grand canonical Monte Carlo simulation devised for a non-equilibrium situation. 
The input variable of the LEMC simulation is the chemical potential profile, $\mu_{i}(\mathbf{r})$, that is not constant for a system out of equilibrium, but a space-dependent quantity.
The output variable is the concentration profile, $c_{i}(\mathbf{r})$. 
Thereby, the LEMC simulation establishes the relation between $c_{i}(\mathbf{r})$ and $\mu_{i}(\mathbf{r})$ necessary to apply the NP equation.
The LEMC method correctly computes volume exclusion and electrostatic correlations between ions, so it is beyond the mean-field level of the PNP theory that is routinely applied for nanopores.

The NP equation is solved iteratively with the LEMC simulations in an iteration procedure that ensures that the continuity equation, $\nabla \cdot \mathbf{j}_{i}(\mathbf{r})=0$, is satisfied  (more detail is found in the Appendix and in the ESI).
The resulting NP+LEMC technique provides a solution for the statistical mechanical problem (e.g., the $c_{i}-\mu_{i}$ relation) on the basis of particle simulations, while it still gives an approximate indirect solution for the dynamical problem through the NP equation.
Direct simulation of ionic transport in the implicit-water framework is commonly done by the Brownian Dynamics method \cite{chung-bj-77-2517-1999,im_bj_2000,berti-jctc-10-2911-2014}.
The advantage of the NP+LEMC technique is that we can easily handle cases where direct sampling of ions crossing the pore is problematic such as the case of the bipolar nanopore, through which currents are small due to the depletion of ions in the oppositely charged zones.

\paragraph*{Questions}

The questions that we pose in this study are the following.
Is the implicit-water model a good approximation to the explicit-water case from the device point of view?
Do the two models provide the same device behavior, namely, have the same transfer function for the device?
The transfer function describes the relation between the control signal (input) and the response (output) that the model gives to the input signal.
Strictly speaking, this is the electrical current as a function of voltage that drives the current. 
We can define transfer functions, however, in a more general way: the electrical current as a function of experimentally tunable parameters such as pore charge and bulk concentrations.

The transfer functions are the same on the two modeling levels because we study the same device with the two models.
The implicit model is justified on the basis of its ability to reproduce the transfer functions given by the detailed model.
The MD model, therefore, is the gold standard in this comparison because it contains all the molecular details (explicit water) that are missing from the implicit-water model.
If we consider the device as a black box model, we are satisfied with just answering this question.
As the dimensions of the devices are getting smaller, however, we want to open the box and look inside to understand how it works on the level of molecules, which is necessary to design new devices.

Even if the explicit- and implicit-water models work similarly on the device-level, opening the box reveals that there are profound differences between the two models due to the presence or absence of explicit water molecules.
In this case, new questions arise.
How can the transfer functions be similar despite the differences?
Specifically, why can water molecules be averaged into an implicit background even in confined spaces? 
Using a reduced model we may be ignoring crucial details.
Looking at too much detail, however, can result in missing the forest for the trees.
Unknown errors from inadequately calibrated parameters of detailed models may produce ambiguous artifacts.
What are the details that cannot be coarse-grained and should be included in the reduced model to reproduce the device function?

\begin{figure*}[t]
\begin{center}
\scalebox{1}{\includegraphics*{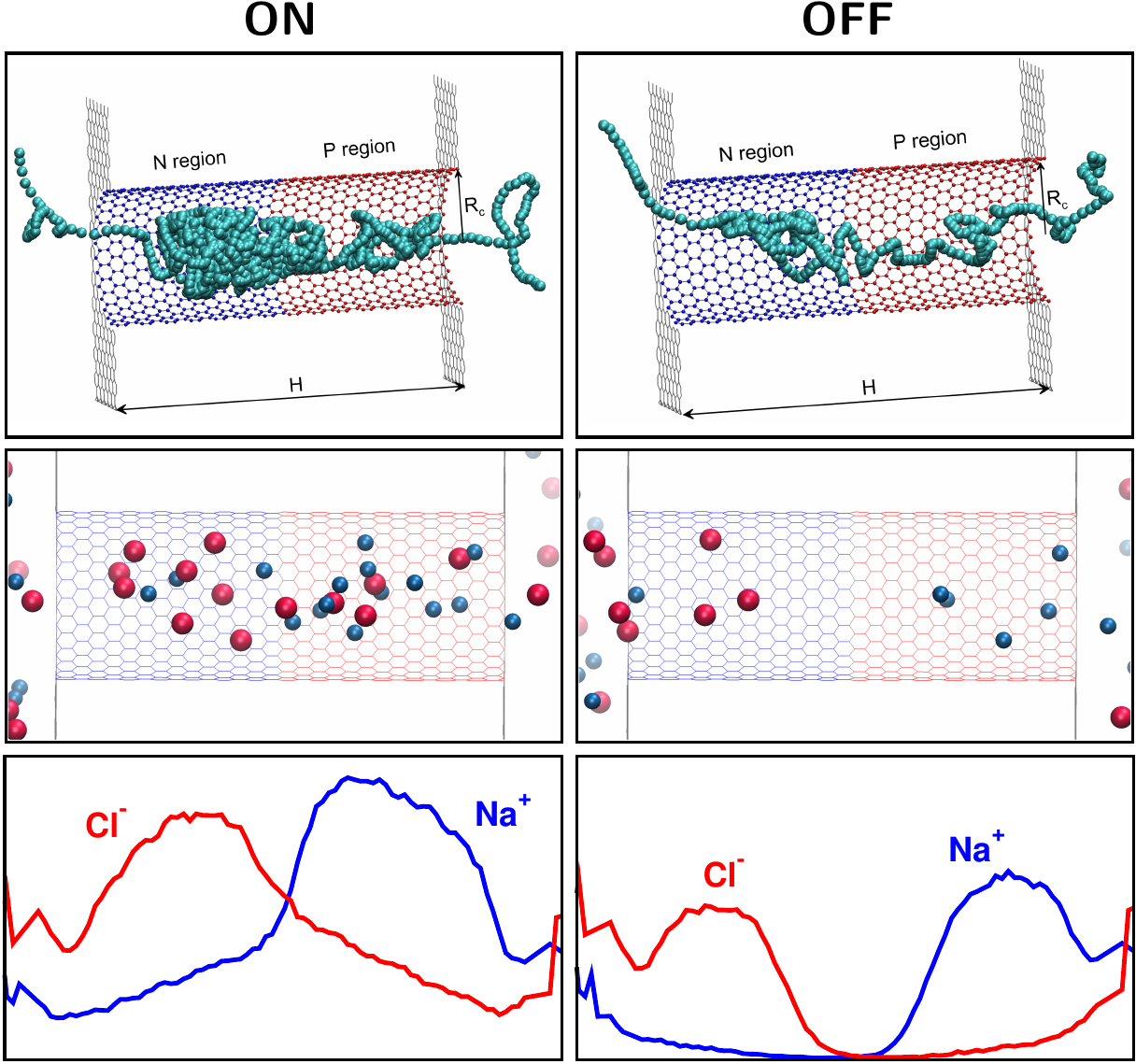}}
\end{center}
\caption{Illustration of the geometry and the rectification mechanism. 
Top panels: Nanopore structure made by the VMD package \cite{vmd}. The pore is formed by a carbon nanotube (CNT) between two carbon nanosheets (CNS) defining the membrane. Distance of CNS atoms in the two sheets (width of membrane) is $H=6.035$ nm, while the distance of CNT atoms from the pore axis (pore radius) is $R_{\mathrm{c}}=1.136$ nm. The pore is made charged by placing partial charges on the C atoms of the CNT. The charges are positive on the left hand side (blue dots, N region) and negative on the right hand side (red dots, P region) of the pore. The value of partial charges is chosen so that the average surface charge density corresponds to a prescribed value, $\pm \sigma$. The series of spheres inside the pore represent trajectories of Cl$^{-}$ ions through the pore plotted in 10 ps time intervals. The Cl$^{-}$ ion was randomly chosen of those that crossed the pore. The trajectories are from simulations for $c=1$ M NaCl and $\sigma = 1 $ $e/$nm$^{2}$ at $\pm 200$ mV voltages (200 mV is ON -- left 
panels, while -200 mV is OFF -- right panels).
Middle panels: randomly chosen snapshots (from a video available in the ESI) with blue and red spheres representing Na$^{+}$ and Cl$^{-}$ ions, respectively. The depletion zone of Na$^{+}$ is formed in the entire N zone, not only in the PN junction in the middle (the same is true for Cl$^{-}$ in the P region). The depletion zones are deeper in the OFF state. 
Bottom panels: illustrative MD concentration profiles (more detailed profiles are shown in Fig.\ \ref{Fig3}). 
}
\label{Fig1}
\end{figure*}

\paragraph*{Bipolar nanopore models}
To address these questions, we have chosen a bipolar nanopore as a test system because it has a very characteristic transfer function: the pore rectifies electrical current.
The output is the current given as a response to the input, the voltage.
The current is markedly different for different signs of the voltage due to the asymmetrical surface charge distribution on the pore wall. 
The sign of the surface charge changes from positive to negative along the central axis of the pore in this study -- we call these N and P regions, respectively, according to the nomenclature of the semiconductor literature (Fig.\ \ref{Fig1}).
In the N zone (with positive surface charge), the anions are the counter-ions and therefore the majority charge carriers, while the cations are the co-ions and the minority charge carriers.
In the P zone, the situation is reversed.
Pore regions with opposite surface charges can be achieved by chemical modifications in the case of PET nanopores by transforming carboxyl groups into amino groups by a coupling agent \cite{vlassiouk_nl_2007}, for example. 
This modification can even be reversible by making the chemical moieties redox-sensitive \cite{ali_sabc_2017}.
The pore charges in the N and P zones do not drive ion tranport; rather, they modulate the concentrations of the various ionic species in the respective zones. 
The driving force of drift-diffusion of ions is the difference of the concentrations and/or applied voltage on the two sides of the membrane \cite{lynden_Bell_jcp_1996,chen_bj_1997,1998_nonner_bj_1287,kurnikova_bj_1999,cardenas_bj_2000,gillespie_jpcm_2002,im_jmb_2002,gillespie-jpcb-109-15598-2005,peter_bj_2005,cervera_epl_2005,cervera_jcp_2006,mamonov_bj_2006,constantin_pre_2007,vlassiouk_nl_2008,gillespie-bj-95-609-2008,he-jacs-131-5194-2009,song_plosone_2011,cervera_ea_2011}.

Fabrication of such synthetic nanopores have become possible recently \cite{siwy_prl_2002,siwy_ss_2003,siwy_nim_2003,howorka_siwy_chapt11_2009,guo_acr_2013,gibb_chapter_2013,guan_nanotech_2014,zhang_nanotoday_2016}. 
Their common property (that distinguishes them from micropores) is that their radii are smaller than (or similar in size to) the characteristic screening length of the electrolyte (NaCl in this study) that fills the channel.
The dimensionality of these systems makes them central building blocks of devices for various applications from DNA sequencing \cite{otto_chapter_2013} through switches \cite{wang_cs_2017}  to chemical and biosensing \cite{sexton_mbs_2007,howorka_csr_2009,piruska_csr_2010,howorka_nbt_2012}.  
Rectification appears (and the nanopore can be called a nanofluidic diode) if the nanopore is asymmetrical in geometry (e.g. conical nanopores) or in charge distribution (e.g. bipolar nanopores studied in this work). 
What makes the bipolar nanopore more suitable to be the test system of our multiscale modeling study is the fact that the mechanism of rectification is mainly electrostatic in nature and that it rectifies even if the dimensions of the pore are small (1 nm radius and 6 nm length in this work) and therefore suitable to study with MD. 

Although bipolar nanopores have been studied with the PNP theory extensively \cite{daiguji_nl_2005,constantin_pre_2007,vlassiouk_nl_2007,karnik_nl_2007,vlassiouk_acsnanno_2008,kalman_am_2008,yan_nl_2009,nguyen_nt_2010,szymczyk_jpcb_2010,singh_jap_2011,singh_jpcb_2011,singh_apl_2011,van_oeffelen_plosone_2015,tajparast_bba_2015}, we are not aware of any paper where this system is examined by molecular simulations apart from our previous study for a rectifying bipolar ion channel \cite{hato-cmp-19-13802-2016} and our parallel study where we compare to PNP \cite{matejczyk-jcp-submitted-2017}. 
We are aware only of a few MD studies \cite{aksimentiev_ieee_2009,cruzchu_fd_2009,cruzchu_jpcc_2009,chen_small_2011,gamble_jpcc_2014,ge_ms_2016} for other types of nanopores.


In this study we pay special care to modeling the nanopore in the MD and the NP+LEMC calculations as similarly as possible, so that the major difference between the two systems is the treatment of water, compared to which other dissimilarities are minor. 
This makes this work a case study for studying the differences between the explicit and implicit water models.
We connect the two modeling levels by estimating the diffusion coefficients of ions in the pore (for the NP+LEMC calculations) from the MD simulations.
We fit the parameters of the $D_{i}(\mathbf{r})$ functions so that NP+LEMC results for the ionic currents reproduce MD results as closely as possible.
In this study, we choose the simplest assumption for the $D_{i}(\mathbf{r})$ profile: it is a piecewise constant function that is different inside the pore and in the bulk: $D_{i}^{\mathrm{pore}}$ and $D_{i}^{\mathrm{bulk}}$. 
The difference of the $D_{i}^{\mathrm{pore}}$ and $D_{i}^{\mathrm{bulk}}$ values accounts for all those effects that are different in the pore and in the bulk including different structure of water molecules around the ions and the presence of the pore (confinement and being trapped in electrostatic wells created by the pore charges).
We fit the $D_{i}^{\mathrm{pore}}$ values for one state, fix them, and check whether they are appropriate in other cases too.

We work with a cylindrical nanopore model whose dimensions are relatively small, only $\sim 1$ nm in radius and $\sim 6$ nm in length. 
This radius is quite close to that attainable in experiments at the tip of conical nanopores. 
The length is far from the experimentally realistic $\mu$m length scale, but we are restricted by the fact that we use particle simulations. 
The small pore radius allows the overlapping of the double layers formed at the pore walls at the 0.5 and 1 M concentrations used in this study (wider pores are considered elsewhere \cite{matejczyk-jcp-submitted-2017}).
Large concentrations were necessary to achieve a reasonable sampling in the MD simulations.
Small pore radius is also useful for the purpose of this study because it amplifies the role of water molecules.

\afterpage{\clearpage}

The surfaces of the pore and the membrane are smooth. 
This is advantageous because we can evaluate the results more easily, we can avoid artifacts from surface roughness, and we can create toy models for both the MD simulations (using a carbon nanotube (CNT) as template, see Fig.\ \ref{Fig1}), and the NP+LEMC calculations (using hard walls, see Fig.\ SI\ 12).
This way, the results of the two methods are comparable and the main difference between them is the treatment of water.

\section*{Results and Discussion}
\label{sec:res}

\subsection*{Results for the device level}

First, we looked at the nanopore from the device perspective, namely, we studied how the response function behaves in the two modeling frameworks.
Because MD sampling is weak for small concentrations and voltages, we restrict ourselves to voltages $\pm 200$ mV and concentration $c=1$ M in the main text.
The effect of the bulk concentration and voltage-dependence (current-voltage curves) is found in the ESI showing mainly NP+LEMC results with a few MD data (Figs.\ SI\,1 and SI\,5).
For a full comparison between the two models, we focus on the dependence of currents on the surface charge density on the nanopore wall, $\sigma$.
We place $\sigma$ surface charge on the left hand side of the pore wall and $-\sigma$ on the right hand side (see Fig.\ \ref{Fig1}), therefore, the value of $\sigma$ characterizes the strength of the polarity of the pore. 
We define rectification as the ratio of the magnitudes of the  ON (200 mV) and the OFF (-200 mV) currents (ground is on the left hand side).

\begin{figure}[t]
 \begin{center}
\scalebox{0.45}{\includegraphics*{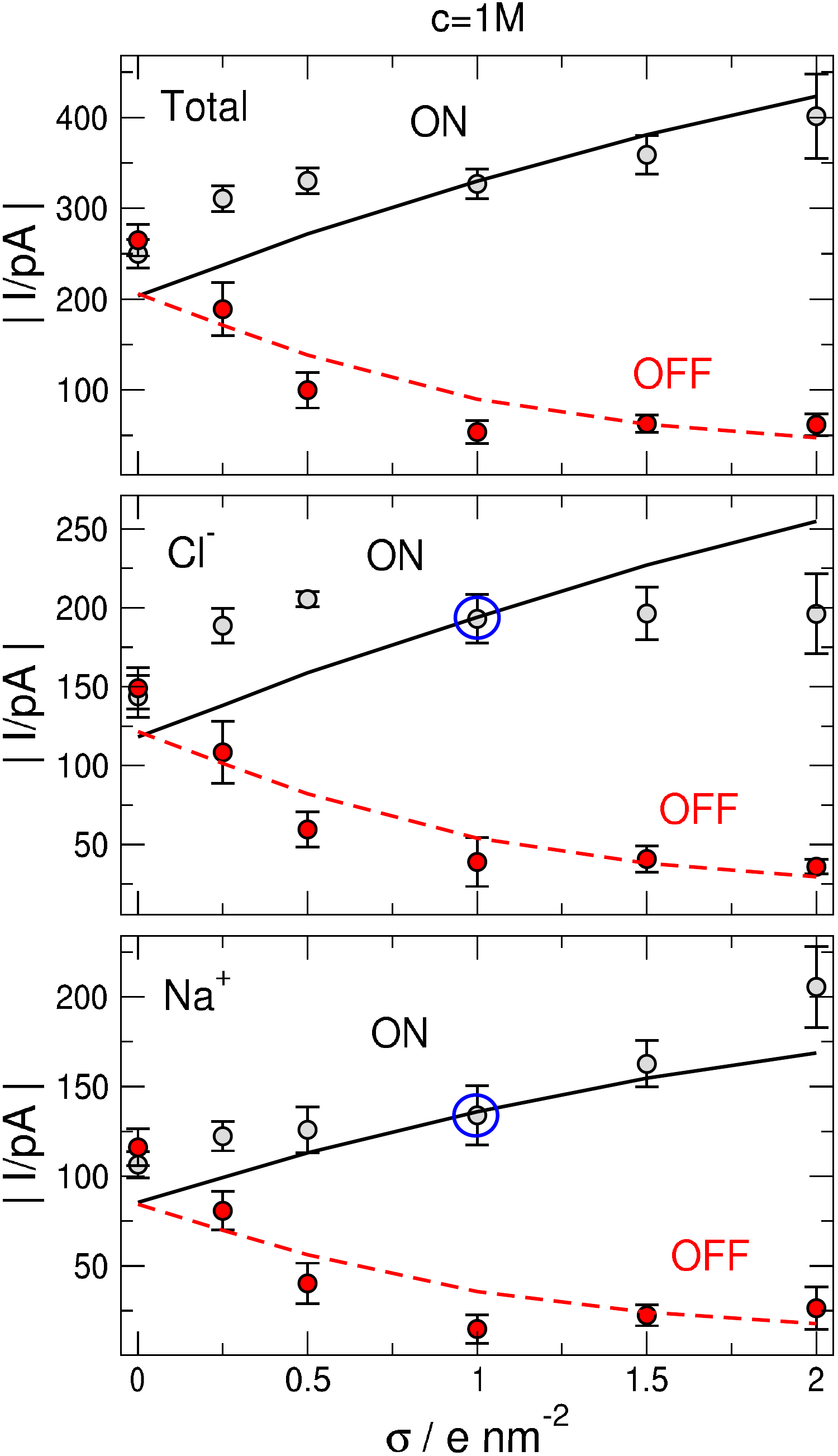}}
 \end{center}
\caption{Currents as functions of surface charge density for $c=1$ M for the ON (200 mV) and OFF (-200 mV) sign of the voltage. Lines and symbols represent NP+LEMC and MD results, respectively. Blue circles indicate the state point at which the diffusion coefficient values, $D_{\mathrm{Na}^{+}}^{\mathrm{pore}}$ and $D_{\mathrm{Cl}^{-}}^{\mathrm{pore}}$, are fitted to the MD currents, therefore, MD and NP+LEMC results coincide in this case.}
 \label{Fig2}
\end{figure}

Figure \ref{Fig2} shows currents as functions of $\sigma$ for bulk concentration $c=1$ M (the analogous figure for $c=0.5$ M is Fig.\ SI\,2 in the ESI).
The MD and NP+LEMC currents coincide for $\sigma=1$ $e$/nm$^{2}$ in the ON state, because we fitted the ionic diffusion coefficients inside the pore, $D_{i}^{\mathrm{pore}}$, so that NP+LEMC results reproduce the MD currents in this case.
We obtained the values $D_{\mathrm{Na}^{+}}^{\mathrm{pore}}=0.3312\times 10^{-9}$m$^{2}$s$^{-1}$ and $D_{\mathrm{Cl}^{-}}^{\mathrm{pore}}=0.485\times 10^{-9}$ m$^{2}$s$^{-1}$ from this adjustment.
Then, we fixed these diffusion coefficient values and never adjusted them again; they remain the same for other surface charges, voltages, and concentrations.

The relevant questions from the point of view of multiscaling is whether the NP+LEMC calculations using these fixed $D_{i}^{\mathrm{pore}}$ values can reproduce the MD data for other cases too, namely, in the OFF state, for other surface charges, and other concentrations.
Posing the question in a different way; are these $D_{i}^{\mathrm{pore}}$ values transferable?
If they are transferable, the reduced model ``knows'' the basic physics necessary to describe the device behavior, so no additional fitting is needed.
This means that the approximations built into the reduced model smear only the ``unimportant'' degrees of freedom into response functions ($\epsilon$ and $D_{i}(\mathbf{r})$), but leave the ``important''  degrees of freedom unaffected. 
A crucial element  of multiscaling is identifying the ``important'' degrees of freedom by doing the calculations at the various resolutions and comparing the results.

Let us examine Fig.\ \ref{Fig2} in more detail.
When the pore is uncharged ($\sigma=0$ $e$/nm$^{2}$), the currents at the positive and negative voltages are the same (apart from statistical uncertainties; see ESI for details), because the pore is symmetric.
When $\sigma$ is increased, the currents in the ON and OFF states increasingly deviate. 
The current (in absolute value) increases in the ON state, while it decreases in the OFF state, i.e., rectification increases with increasing $\sigma$.
The MD and NP+LEMC curves are in qualitative agreement regarding this trend.
This implies that the $D_{i}(z)$ profiles as constructed in this work (experimental value in the bulk, $D_{i}^{\mathrm{bulk}}$, while a constant, fitted, and fixed value in the pore, $D_{i}^{\mathrm{pore}}$) modulate current in a way that agrees with the explicit-water MD model.

\begin{figure*}[t]
 \begin{center}
\scalebox{0.55}{\includegraphics*{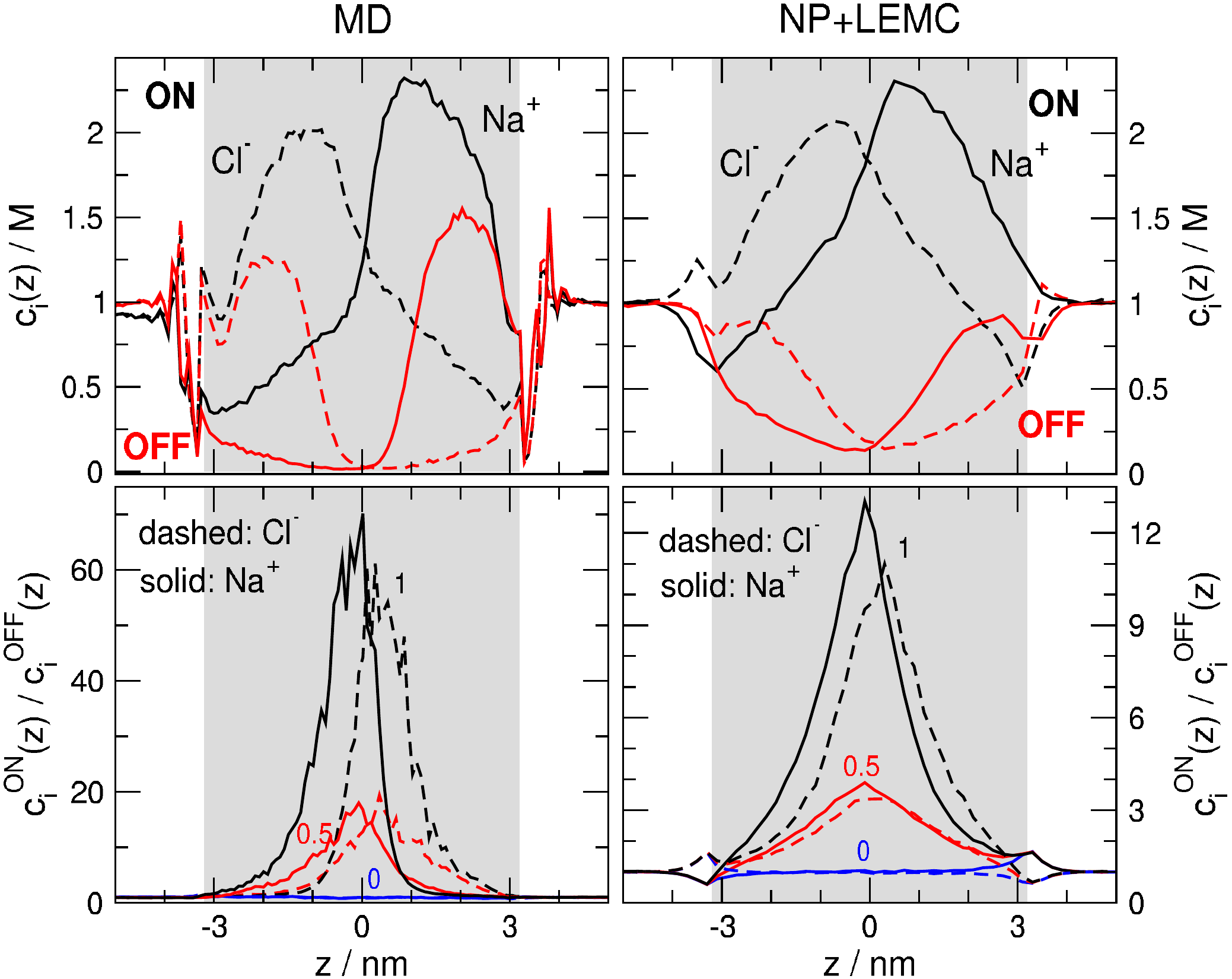}}
 \end{center}
\caption{Concentration profiles as obtained from MD (left panels) and NP+LEMC (right panels). In the top panels, the Na$^{+}$ (solid lines) and Cl$^{-}$ (dashed lines) profiles are shown for the ON (black lines) and OFF (red lines) signs of the voltage for $c=1$ M and $\sigma =1$ $e$/nm$^{2}$. 
In the bottom panels, the ratio of concentration profiles in the ON and OFF states are shown for various values of $\sigma$ (see the numbers near the curves).
Concentration profiles have been computed in the same way in the two models: the average number of ions in a slab has been divided by the effective volume available for the ions. Inside the pore, the same cross section was used  (radius 1 nm) in both cases.}
\label{Fig3}
\end{figure*}

If we want to look into the black box, we need to analyze space-dependent concentration and potential profiles.
From the point of view of current, which is the integral of the flux density over the cross section of the pore, the $z$-dependent concentration profiles (averages over cross sections) are the most relevant profiles.
They describe which ion is present in which region (N and P) of the nanopore and are the first-order determinants of the current.

\begin{figure*}[t]
\begin{center}
\scalebox{1.1}{\includegraphics*{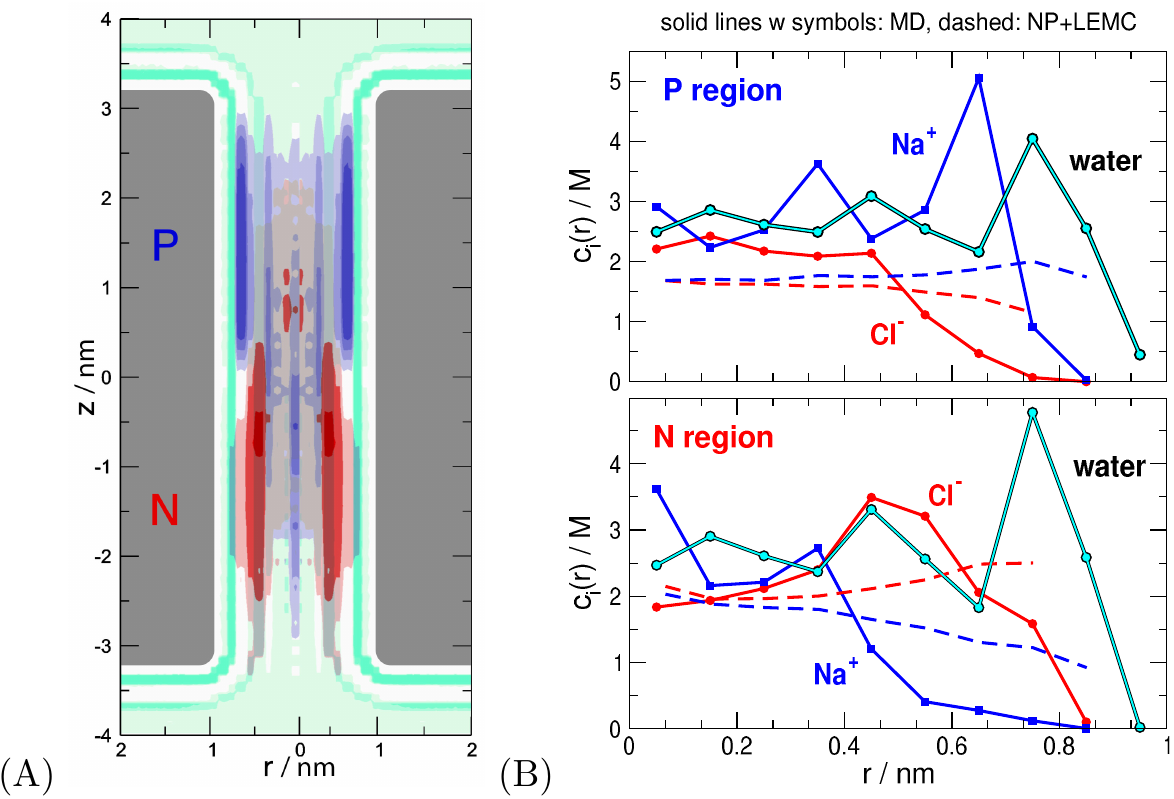}}
\end{center}
\caption{(A) Structure of water molecules (turquoise), Na$^{+}$ (blue), and Cl$^{-}$ (red) ions as obtained from MD simulations in the ON state for $\sigma=1$ $e$/nm$^{2}$ and $c=1$ M.
Coloring is designed to overemphasize peaks in order to show structures. 
(B) Radial profiles by averaging over the P region ($0<z<3$ nm, top panel) and the N region ($-3<z<0$ nm, bottom panel) are plotted with symbols connected with lines (same colors as in panel A). Water profiles are divided by 20 to make them comparable to ion profiles.
The curves obtained from the implicit-water model are also shown (dashed lines).}
\label{Fig4}
\end{figure*}

The top panels of Fig.\ \ref{Fig3} show concentration profiles obtained from MD (left panels) and NP+LEMC (right panels) calculations.
The two models describe the mechanism of rectification in qualitative agreement.
\emph{As a primary effect}, the pore charges produce depletion zones for the respective ions, e.g., the positive pore charges on the left hand side (N zone) result in depletion zones in the Na$^{+}$ profiles (the same is true for the Cl$^{-}$ profiles in the P zone).
The Na$^{+}$ ions, therefore are co-ions in the N zones, while the Cl$^{-}$ ions are counter-ions there (reversed in the P zone).
In this system, diffusion is limited for both ionic species by their respective depletion zones because these zones are the largest resistance elements if we imagine the ionic pathway as resistors connected in series in an equivalent circuit; if an ion species is not there, it cannot conduct.
Depletion zones, therefore, are more important than peaks.

\emph{As a secondary effect}, the applied field modulates the effect of pore charges. 
It increases the concentrations (of both ions) in the ON state, while it decreases them in the OFF state.
Depletion zones of the ions, therefore, are deeper in the OFF state than in the ON state.
This is the basic reason of rectification.
It is important to stress, however, that the peaks are also lower in the OFF state.
The depletion zones are not independent from the peaks. 
Co-ions and counter-ions are strongly correlated electrostatically, therefore, co-ions are present in their depletion zones because the counter-ions are present there.

As far as the $\sigma$-dependence is concerned, ON concentrations increase, while OFF concentrations decrease with increasing $\sigma$ as shown by the bottom panels of Fig.\ \ref{Fig3}, where the ratio of the ON and OFF concentrations is shown for different $\sigma$ values (the concentration profiles themselves are in Fig.\ SI\ 4).
The ratio increases with increasing $\sigma$. 
The ratio is larger in the case of the MD data (corresponding to deeper depletion zones), which is reflected in the smaller OFF currents compared to the NP+LEMC results. 
Better agreement between MD and NP+LEMC could be achieved by adjusting different diffusion constant values in the N and P zones (different mobilities in the N and P zones are implied by the top panels of Fig.\ \ref{Fig1}), but this is beyond the scope of this study.

The results obtained for the operation of the nanopore as a device (and the accompanying $c_{i}(z)$ profiles) are quite similar on the two modeling levels.
This is surprising given that the treatments of water in the two models are so profoundly different. 
Consequences of the differences can be revealed if we open the black box even wider and dive into details beyond the cross-sectionally averaged $c_{i}(z)$ profiles.

\subsection*{Results for the molecular level}
The presence of water molecules and their effect on ions can be seen by plotting the full $c_{i}(z,r)$ profiles.
Figure \ref{Fig4}A shows contour plots for the ON state as obtained from MD simulations (the plot for the OFF state is seen in Fig.\ SI\ 7).
A clear layering structure as a function of the $r$ coordinate  is present.
There are two distinct peaks of water, one is near the pore wall solvating the pore charges. 
The ions' behavior depends on the region in which we observe them.
Cl$^{-}$ ions have a large peak just ``behind'' the solvating water layer in the N region, while they rather accumulate in the pore center in the P region.
Na$^{+}$ ions, on the other hand, show the opposite behavior, exhibiting two peaks in the P region.

\begin{figure*}[t]
 \begin{center}
\scalebox{1}{\includegraphics*{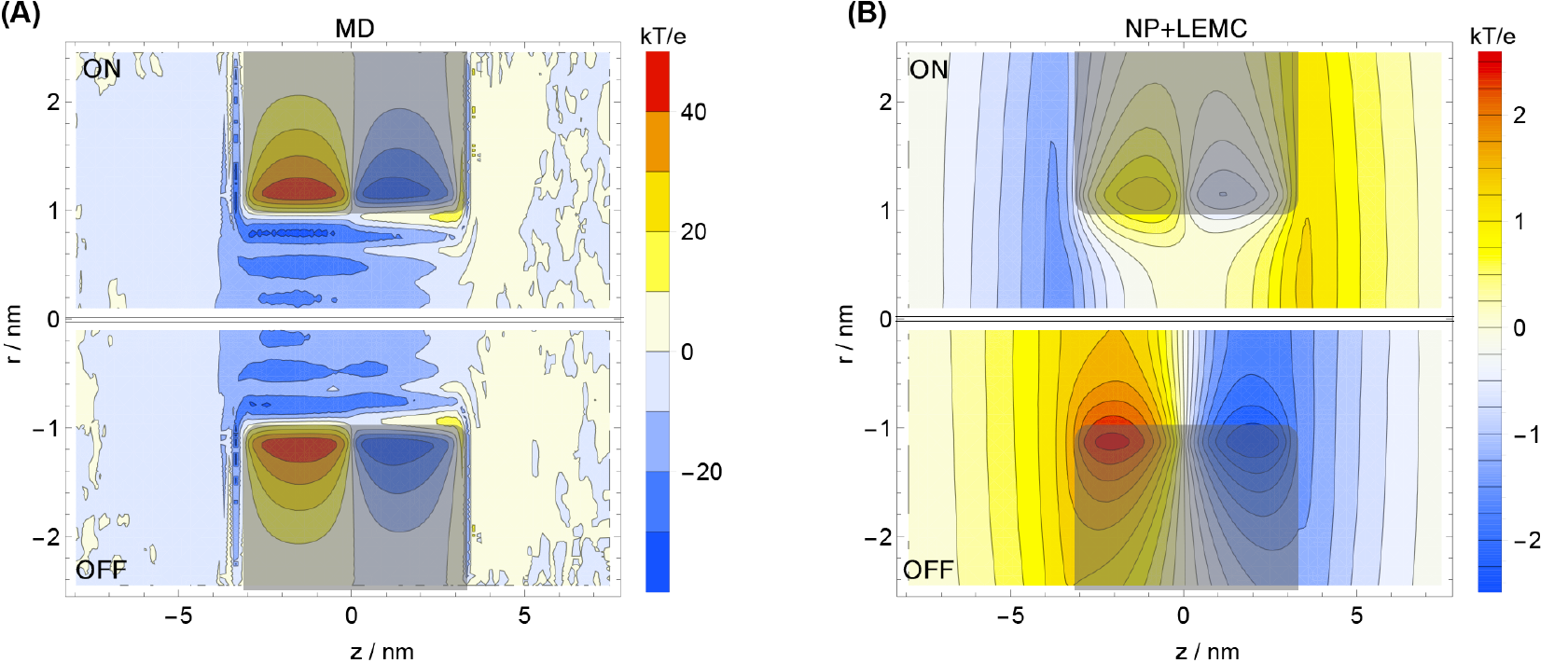}}
 \end{center}
\caption{Electrical potential profiles as obtained from MD (left panel) and NP+LEMC (right panel) simulations. The profiles do not include the applied potential. The MD profiles include the contributions of fixed charges (on the pore wall), ions, and water. The MC profiles include only the effect of fixed charges and ions divided by $\epsilon$.
Results are shown for the ON (top halves) and OFF (bottom halves) signs of the voltage. 
The profiles have been computed from the average charge profiles provided by the simulations from solving Poisson's equation (see Appendix and Fig.\ SI\ 8).
}
\label{Fig5}
\end{figure*}

These effects are better (and more quantitatively) observed in Fig.\ \ref{Fig4}B, where the $c_{i}(r)$ profiles are plotted averaged over the $z$-coordinate in the N region (top panel) and in the P region (bottom panel).
Here, the NP+LEMC profiles for Na$^{+}$ and Cl$^{-}$ are also shown.
Although they are unstructured, they reproduce the relative quantities of the two ion species in the respective regions.

Therefore, despite the differences in the radial structure, the MD and NP+LEMC ionic profiles show basically the same behavior in the $z$ dimension.
This is the behavior that is relevant for the calculation of the current, and therefore, for the function of the device.
The $r$-dependence seems to belong to the class of the ``unimportant'' degrees of freedom vis-\`{a}-vis ion current, because its accurate reproduction is not necessary to properly model the operation of the nanopore as a rectifying diode. 
The $z$-dependence, on the other hand, is crucially important.
This explains why 1-dimensional PNP calculations work so well for nanopores.  
The water molecules then also seem to belong to the class of ``unimportant'' degrees of freedom, in the sense that they do not contribute to the current (they have no net charge) and do not affect the axial $c_{i}(z)$ profiles of ions substantially. 
Also, the treatment of their dynamic effect on the ions as a friction via the diffusion coefficients in the NP equation seems to be a sufficient approximation.

Their other effect on ions, screening, also works very differently in the implicit and explicit-water models.
To perceive these profoundly different ways of screening as given by MD and NP+LEMC simulations, we consider the electrical potential profiles produced by the various species in the two models. 
The electrostatic potential from water is present in the MD simulations explicitly and provides screening of the potential from ions (and pore charges) in an additive way. 
In the NP+LEMC simulations, screening is provided by polarization charges $(1/\epsilon-1)q$ that are right on top of the $q$ ionic point charges. 
In effect, this correponds to just damping the electric field of ions by dividing it by $\epsilon$ (see Eq.\ \ref{eq:pm}).

Figure \ref{Fig5} shows potential profiles as contour plots over the $(z,r)$ coordinates obtained from MD (left panel) and NP+LEMC (right panel) at the same conditions.
These potentials do not contain the applied potential, only those produced by free charges in the system: ions, pore charges, and water (if present). 
Top halves are for the ON state, while bottom halves are for the OFF state.
The differences between the two models are striking.

In the implicit-water model (Fig.\ \ref{Fig5}B), only ions can respond to changes in the electric field.
The $(1/\epsilon -1)q$ polarization charges move together with the ions so they cannot respond to the electric field independently.
The effect of the applied field, therefore, is clearly visible.
More ions are available for screening in the ON state, so the potential is smaller (in absolute value) in the ON state.
In the OFF state, due to the low concentration of ions, the pore charges dominate the potential.

In the case of the explicit-water model (Fig.\ \ref{Fig5}A), on the other hand, water molecules also produce a counter-field to external effects.
The first-order external effects to which both ions and water respond are the applied field and pore charges (these are fixed).
Water molecules also respond to the field of ions (and vice versa, of course).
Because ions move slowly and they alone (without water) form a low-density (dilute) system, the characteristic speed at which they form the ionic distribution is relatively small.
Water molecules, on the other hand, are high-density (liquid) and rotate quickly, so they accomodate themselves to the movement of ions and other external fields fast.
The electrolyte, therefore, is a conducting material in which mobile charges are found in high density that respond to external fields and exert a counter-field that produce close to constant potential in the electrolyte.

\begin{figure*}[t]
 \begin{center}
\scalebox{1}{\includegraphics*{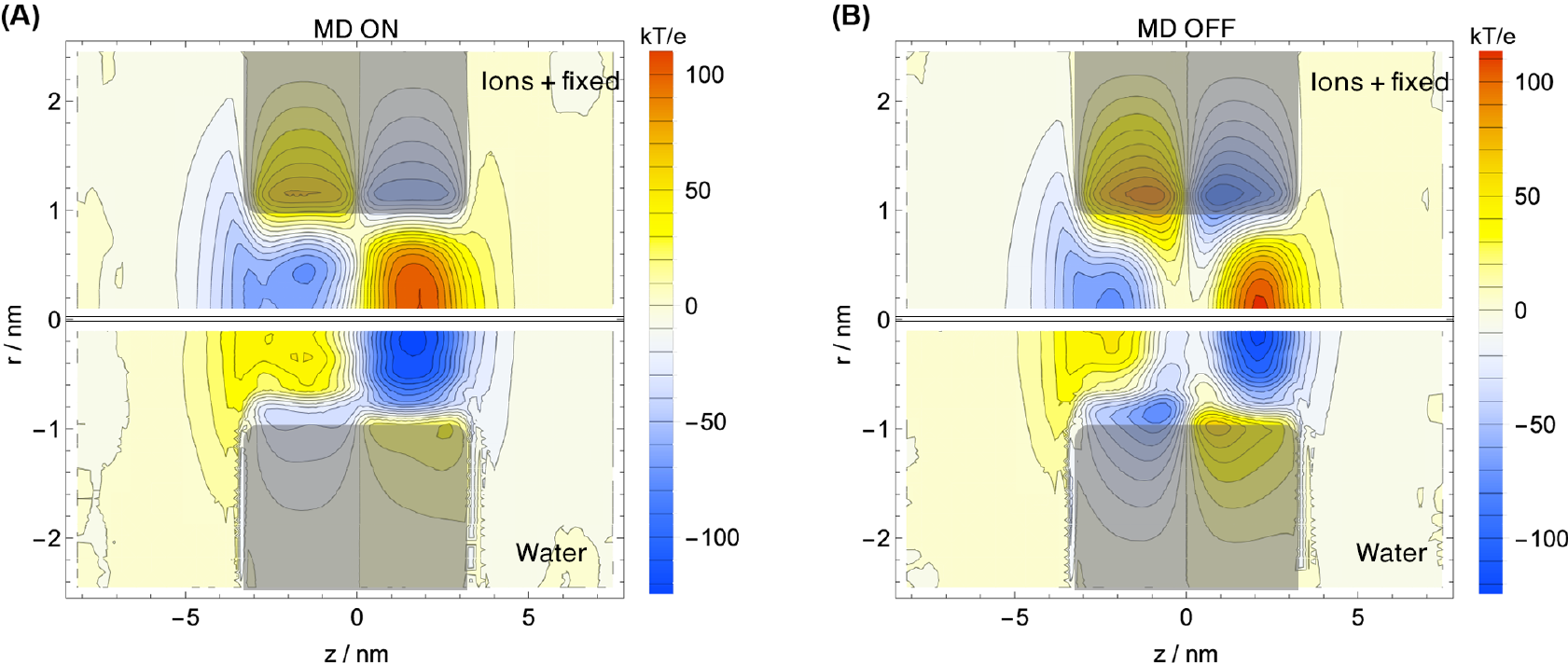}}
 \end{center}
\caption{The two components of the electrical potential of free charges as obtained from MD: potential of the ions and the fixed pore charges (top halves) and potential of water (bottom halves). Results are shown for the ON (panel A) and OFF (panel B) signs of the voltage.}
\label{Fig6}
\end{figure*}

Because the potentials in the implicit-water model are uniformly damped by $\epsilon=78.5$, the potential varies in a narrow range ($-2.5<e\Phi (z,r)/kT<2.5$).
In the explicit-water case, on the other hand, the total potential changes in a much wider range (Fig.\ \ref{Fig5}B vs.\ A).
For this reason, the effect of the relatively small applied field ($\pm 200$ mV) is hardly visible.

The effect of the applied field, on the other hand, is better visible on the potentials produced by the ions (together with pore charges) and water molecules separately.
These potentials are plotted in Fig.\ \ref{Fig6}.
Figures \ref{Fig6}A and B show the results for the ON and OFF states, respectively.
Water molecules produce an electrical potential (bottom halves) that has an opposite sign compared to the potential of ions and pore charges (top halves).

Differences in the potentials of ions and pore charges in the ON and OFF states are consequences of the differences in the ionic profiles in the ON and OFF states.
Water molecules respond easily to the changes in ion distributions (as described above), so their potentials are also different in the ON and OFF states.

Another important difference between the explicit- and implicit-water models is whether ionic double layers are formed near the membrane on the two sides of it.
In the implicit-water case they are shown by the splitting cation and anion profiles (Fig.\ \ref{Fig3} and Fig.\ SI\ 9) and by the gradually rising/declining potential profiles (Fig.\ \ref{Fig5}B and Fig.\ SI\ 10) on the two sides of the membrane. 
The signs of the double layer potentials are the opposite in the ON and OFF states and their formation is closely correlated with the relative quantities of counter-ions vs.\ co-ions in the pore region adjacent to the double layer (see our other papers \cite{hato-cmp-19-13802-2016,matejczyk-jcp-submitted-2017} for details). 
The presence of the double layers, however, is not necessary to produce rectification.
They are absent in the explicit-water MD simulations, still, the different currents in the ON and OFF states are present.

\section*{Conclusions}

We found definite differences between the results of the explicit- and implicit-water models with respect to (1) how they describe the radial dependence of the particle profiles, (2) whether the double layers on the two sides of the membrane are formed or not, and (3) how the screening of ionic and pore charges by water is done.
Still, the $z$-dependence of the ionic concentration profiles is similar in the two models.
The effect of the pore charges producing the depletion zones and the effect of the applied field in modulating the effect of pore charges are so powerful that ions respond to them in a similar way no matter whether the water molecules are there explicitly or their screening effect is just a scaling factor in the denominator of the Coulomb potential.

While this is a simple case study, it does answer the paradox of why reduced models work (i.e., reproduce experimental results) in regimes (e.g., confined geometries) where they ought not to work because their molecular-level physics is too approximate.
Our results show that some atomic details do not matter for device input-output properties, and that reduced models that reproduce experiments get the overall device physics right, even though they may get some aspects of the molecular-scale physics wrong.
That is, reduced models show what device-level physics is necessary/sufficient to predict device characteristics and to design new devices.
The results of reduced models should not be overanalyzed, however.
The nanoscale-level results they produce are likely to be incorrect in many ways and MD simulations should be used if these details are desired.

For the foreseeable future, however, fully atomistic simulations will probably remain too computationally expensive to regularly predict experiments, and reduced models will remain the only way to compare with experiments in many cases.
Our work provides computational insight into how these reduced models should be interpreted.

 \section*{Appendix}
 \label{sec:modelsandmethods}

All-atom MD simulations were performed with the GROMACS (v.5.0.4) program suite \cite{Berendsen199543,Pronk01042013} using the leap-frog integrator with a 2 fs time step. 
The CHARMM27 force field \cite{charmm} was implemented that included $\sim$11,000 SPC water molecules, 110-190 Lennard-Jones (LJ) and point charge type anions/cations (depending on the concentration and pore charge density), and LJ atoms for the CNT and CNS. 
The dimensions of the simulation box were 6$\times$5.2$\times$16.8 nm, with periodic boundary conditions (PBC) applied in all spatial directions. 
The systems were thermostated to 298.15 K by a modified version of the Berendsen (velocity rescaling) algorithm \cite{bussi_jcp_2007}.
An ion was considered to cross the channel if it is initially at one side of the membrane and then ends at the opposite side of the membrane after propagating through the channel.

In the reduced model designed for NP+LEMC calculations the membrane and the pore penetrating it are defined by hard walls with which the ions cannot overlap.
The cylindrical pore's radius and length were calculated to mimic the CNT model of the MD simulations as closely as possible on the basis of an estimated distance of closest approach of ions to the carbon atoms.
We used the values $R=0.97$ nm and $H=6.4$ nm for the pore radius and length, respectively.
The fractional point charges have been placed at the same positions as in the CNT model. 
The Na$^{+}$ and Cl$^{-}$ ions are charged hard spheres with radii 0.095 and 0.181 nm) immersed in a dielectric continuum ($\epsilon =78.5$) that models the solvent implicitly (Eq.\ \ref{eq:pm}).
The bulk diffusion coefficients of Na$^{+}$ and Cl$^{-}$ were $D_{\mathrm{Na}^{+}}^{\mathrm{bulk}}=1.333\times 10^{-9}$m$^{2}$s$^{-1}$ and $D_{\mathrm{Cl}^{-}}^{\mathrm{bulk}}=2.032\times 10^{-9}$ m$^{2}$s$^{-1}$, respectively.
The problem is solved on a discretized grid iteratively; the electrochemical potential is changed until the continuity equation, $\nabla \cdot \mathbf{j}_{i}(\mathbf{r})=0$, is satisfied.
The resulting profiles fluctuate around limiting distributions due to statistical uncertainties in the LEMC simulations.
The final results are obtained as running averages.
Computational time is measured in days (sometimes, weeks) for MD simulations, in hours for the NP+LEMC simulations, but only in seconds for PNP calculations.

Boundary conditions are treated differently in MD and NP+LEMC. 
In MD, a $\pm 0.012$ V/nm homogeneous external electric field was applied along the $z$-direction to achieve $\pm 200$ mV potential difference.
Ions leaving the cell at one end are fed back on the other end due to PBC applied in the $z$ dimension.
In NP+LEMC, the simulation cell is a finite cylinder due to the rotational symmetry \cite{boda-jctc-8-824-2012,boda-jml-189-100-2014,boda-arcc-2014}.
Two cylindrical compartments on the two sides of the membrane represent the two bulk regions between which the ion transport flows.
Both concentrations and electrical potentials are prescribed on the half-cylinders confining these bulk regions.

The potential profiles can be computed either (1) by inserting test charges and sampling the potential on the fly, or (2) by solving Poisson's equation for the averaged $(z,r)$-dependent charge profiles obtained from the simulations (with Dirichlet boundary conditions of assuming zero potential on the closed surface of the confining cylinder).
Agremeent between the two methods was excellent for the NP+LEMC data, while it was reasonable for the MD data (see Fig.\ SI\ 8).
The potential profiles shown in Figs.\ \ref{Fig5} and \ref{Fig6} have been obtained from the second method.


\section*{Acknowledgements}
\label{sec:ack}

We gratefully acknowledge the financial support of the Hungarian National Research Fund (OTKA NN113527) in the framework of ERA Chemistry.
This material is based upon work supported by the National Science Foundation under Grant No. 1402897 (to D.G.).

%

\begin{thebibliography}{10}

\bibitem{cervera_jcp_2006}
J.~Cervera, B.~Schiedt, R.~Neumann, S.~Mafe, and P.~Ramirez.
\newblock Ionic conduction, rectification, and selectivity in single conical
  nanopores.
\newblock {\em J. Chem. Phys.}, 124(10):104706, 2006.

\bibitem{kalman_am_2008}
E.~B. Kalman, I.~Vlassiouk, and Z.~S. Siwy.
\newblock Nanofluidic bipolar transistors.
\newblock {\em Adv. Mater.}, 20(2):293--297, 2008.

\bibitem{cheng_acsnano_2009}
L.-J. Cheng and L.~J. Guo.
\newblock Ionic current rectification, breakdown, and switching in
  heterogeneous oxide nanofluidic devices.
\newblock {\em ACS Nano}, 3(3):575--584, 2009.

\bibitem{hoffmann_l_2013}
Jordan Hoffmann and Dirk Gillespie.
\newblock Ion correlations in nanofluidic channels: {Effects} of ion size,
  valence, and concentration on voltage- and pressure-driven currents.
\newblock {\em Langmuir}, 29(4):1303--1317, 2013.

\bibitem{pietschmann_pccp_2013}
J.-F. Pietschmann, M.-T. Wolfram, M.~Burger, C.~Trautmann, G.~Nguyen,
  M.~Pevarnik, V.~Bayer, and Z.~Siwy.
\newblock Rectification properties of conically shaped nanopores: consequences
  of miniaturization.
\newblock {\em Phys. Chem. Chem. Phys.}, 15:16917--16926, 2013.

\bibitem{perezmitta_pcp_2016}
G.~P\'{e}rez-Mitta, A.~Albesa, M.~E. Toimil-Molares, C.~Trautmann, and
  O.~Azzaroni.
\newblock The influence of divalent anions on the rectification properties of
  nanofluidic diodes: {Insights} from experiments and theoretical simulations.
\newblock {\em ChemPhysChem}, 17(17):2718--2725, 2016.

\bibitem{2000_nonner_bj_1976}
W.~Nonner, L.~Catacuzzeno, and B.~Eisenberg.
\newblock {Binding and selectivity in {L}-type calcium channels: {A} mean
  spherical approximation.}
\newblock {\em Biophys. J.}, 79(4):1976--1992, 2000.

\bibitem{gillespie-bj-95-2658-2008}
D.~Gillespie and D.~Boda.
\newblock The anomalous mole fraction effect in calcium channels: {A} measure
  of preferential selectivity.
\newblock {\em Biophys. J.}, 95(6):2658--2672, 2008.

\bibitem{gillespie-bj-2008}
D.~Gillespie.
\newblock {Energetics of Divalent Selectivity in a Calcium Channel: {The}
  Ryanodine Receptor Case Study}.
\newblock {\em Biophys. J.}, 94(4):1169--1184, 2008.

\bibitem{boda-jgp-133-497-2009}
D.~Boda, M.~Valisk{\'o}, D.~Henderson, B.~Eisenberg, D.~Gillespie, and
  W.~Nonner.
\newblock Ion selectivity in {L-type} calcium channels by electrostatics and
  hard-core repulsion.
\newblock {\em J. Gen. Physiol.}, 133(5):497--509, 2009.

\bibitem{dirk-janhavi-mike}
D.~Gillespie, J.~Giri, and M.~Fill.
\newblock {Reinterpreting the anomalous mole fraction effect: {The} {Ryanodine}
  receptor case study}.
\newblock {\em Biophys. J.}, 97(8):2212--2221, 2009.

\bibitem{malasics-bba-1798-2013-2010}
M.~Malasics, D.~Boda, M.~Valisk{\'o}, D.~Henderson, and D.~Gillespie.
\newblock Simulations of calcium channel block by trivalent ions: {Gd$^{3+}$}
  competes with permeant ions for the selectivity filter.
\newblock {\em Biochim. et Biophys. Acta - Biomembranes}, 1798(11):2013--2021,
  2010.

\bibitem{gillespie_bj_2014}
D.~Gillespie, L.~Xu, and G.~Meissner.
\newblock Selecting ions by size in a calcium channel: {The} {Ryanodine
  Receptor} case study.
\newblock {\em Biophys. J.}, 107(10):2263--2273, 2014.

\bibitem{boda-jpcc-118-700-2014}
D.~Boda, \'E. Cs\'anyi, D.~Gillespie, and T.~Krist\'of.
\newblock Dynamic {Monte Carlo} simulation of coupled transport through a
  narrow multiply-occupied pore.
\newblock {\em J. Phys. Chem. C}, 118(1):700--707, 2014.

\bibitem{gillespie-bj-95-609-2008}
D.~Gillespie, D.~Boda, Y.~He, P.~Apel, and Z.S. Siwy.
\newblock Synthetic nanopores as a test case for ion channel theories: {The}
  anomalous mole fraction effect without single filing.
\newblock {\em Biophys. J.}, 95(2):609--619, 2008.

\bibitem{he-jacs-131-5194-2009}
Y.~He, D.~Gillespie, D.~Boda, I.~Vlassiouk, R.~S. Eisenberg, and Z.~S. Siwy.
\newblock Tuning transport properties of nanofluidic devices with local charge
  inversion.
\newblock {\em JACS}, 131(14):5194--5202, 2009.

\bibitem{boda-jctc-8-824-2012}
D.~Boda and D.~Gillespie.
\newblock Steady state electrodiffusion from the {Nernst-Planck} equation
  coupled to {Local Equilibrium Monte Carlo} simulations.
\newblock {\em J. Chem. Theor. Comput.}, 8(3):824--829, 2012.

\bibitem{boda-jml-189-100-2014}
D.~Boda, R.~Kov\'acs, D.~Gillespie, and T.~Krist\'of.
\newblock Selective transport through a model calcium channel studied by
  {Local} {Equilibrium} {Monte} {Carlo} simulations coupled to the
  {Nernst}-{Planck} equation.
\newblock {\em J. Mol. Liq.}, 189:100--112, 2014.

\bibitem{boda-arcc-2014}
D.~Boda.
\newblock Chapter five - monte carlo simulation of electrolyte solutions in
  biology: In and out of equilibrium.
\newblock In Ralph~A. Wheeler, editor, {\em Ann. Rep. Comp. Chem.}, volume~10,
  chapter~5, pages 127--164. Elsevier, 2014.

\bibitem{chung-bj-77-2517-1999}
S.~H. Chung, T.~W. Allen, M.~Hoyles, and S.~Kuyucak.
\newblock {Permeation Of Ions Across The Potassium Channel: {Brownian} Dynamics
  Studies}.
\newblock {\em Biophys. J.}, 77(5):2517--2533, 1999.

\bibitem{im_bj_2000}
W.~Im, S.~Seefeld, and B.~Roux.
\newblock {A Grand Canonical {Monte Carlo}-{Brownian} Dynamics Algorithm For
  Simulating Ion Channels}.
\newblock {\em Biophys. J}, 79(2):788--801, 2000.

\bibitem{berti-jctc-10-2911-2014}
C.~Berti, S.~Furini, D.~Gillespie, D.~Boda, R.~S. Eisenberg, E.~Sangiorgi, and
  C.~Fiegna.
\newblock A {3-D Brownian Dynamics} simulator for the study of ion permeation
  through membrane pores.
\newblock {\em J. Chem. Theor. Comput.}, 10(8):2911--2926, 2014.

\bibitem{vlassiouk_nl_2007}
I.~Vlassiouk and Z.~S. Siwy.
\newblock Nanofluidic diode.
\newblock {\em Nano Lett.}, 7(3):552--556, 2007.

\bibitem{ali_sabc_2017}
M.~Ali, I.~Ahmed, P.~Ramirez, S.~Nasir, S.~Mafe, C.~M. Niemeyer, and
  W.~Ensinger.
\newblock A redox-sensitive nanofluidic diode based on nicotinamide-modified
  asymmetric nanopores.
\newblock {\em Sensors and Actuators B: Chemical}, 240:895--902, 2017.

\bibitem{lynden_Bell_jcp_1996}
R.~M. Lynden-Bell and J.~C. Rasaiah.
\newblock Mobility and solvation of ions in channels.
\newblock {\em J. Chem. Phys.}, 105(20):9266--9280, 1996.

\bibitem{chen_bj_1997}
D.~Chen, J.~Lear, and B.~Eisenberg.
\newblock Permeation through an open channel: {Poisson-Nernst-Planck} theory of
  a synthetic ionic channel.
\newblock {\em Biophys. J.}, 72(1):97--116, 1997.

\bibitem{1998_nonner_bj_1287}
W.~Nonner and B.~Eisenberg.
\newblock {Ion Permeation and Glutamate Residues Linked by
  {P}oisson-{N}ernst-{P}lanck Theory in {L}-type Calcium Channels.}
\newblock {\em Biophys. J.}, 75:1287--1305, 1998.

\bibitem{kurnikova_bj_1999}
M.~G. Kurnikova, R.~D. Coalson, P.~Graf, and A.~Nitzan.
\newblock A lattice relaxation algorithm for three-dimensional
  {Poisson-Nernst-Planck} theory with application to ion transport through the
  {Gramicidin A}channel.
\newblock {\em Biophys. J.}, 76(2):642--656, 1999.

\bibitem{cardenas_bj_2000}
A.~E. C\'{a}rdenas, R.~D. Coalson, and M.~G. Kurnikova.
\newblock Three-dimensional {Poisson-Nernst-Planck} theory studies: Influence
  of membrane electrostatics on {Gramicidin A} channel conductance.
\newblock {\em Biophys. J.}, 79(1):80--93, 2000.

\bibitem{gillespie_jpcm_2002}
D.~Gillespie, W.~Nonner, and R.~S. Eisenberg.
\newblock {Coupling {Poisson-Nernst-Planck} and density functional theory to
  calculate ion flux}.
\newblock {\em J. Phys.: Cond. Matt.}, 14(46):12129--12145, 2002.

\bibitem{im_jmb_2002}
W.~Im and B.~Roux.
\newblock Ions and counterions in a biological channel: {A} molecular dynamics
  simulation of {OmpF} porin from {Escherichia} coli in an explicit membrane
  with 1 {M} {KCl} aqueous salt solution.
\newblock {\em J. Mol. Biol.}, 319(5):1177--1197, 2002.

\bibitem{gillespie-jpcb-109-15598-2005}
D.~Gillespie, L.~Xu, Y.~Wang, and G.~Meissner.
\newblock {({De})constructing the ryanodine receptor: {Modeling} ion permeation
  and selectivity of the calcium release channel}.
\newblock {\em J. Phys. Chem. B}, 109(32):15598--15610, 2005.

\bibitem{peter_bj_2005}
C.~Peter and G.~Hummer.
\newblock Ion transport through membrane-spanning nanopores studied by
  molecular dynamics simulations and continuum electrostatics calculations.
\newblock {\em Biophys. J.}, 89(4):2222--2234, 2005.

\bibitem{cervera_epl_2005}
J.~Cervera, B.~Schiedt, and P.~Ram{\'{\i}}rez.
\newblock A {Poisson/Nernst-Planck} model for ionic transport through synthetic
  conical nanopores.
\newblock {\em Europhys. Lett.}, 71(1):35--41, jul 2005.

\bibitem{mamonov_bj_2006}
A.~B. Mamonov, M.~G. Kurnikova, and R.~D. Coalson.
\newblock Diffusion constant of k$^+$ inside {Gramicidin A}: {A} comparative
  study of four computational methods.
\newblock {\em Biophys. Chem.}, 124(3):268--278, 2006.

\bibitem{constantin_pre_2007}
D.~Constantin and Z.~S. Siwy.
\newblock {Poisson-Nernst-Planck} model of ion current rectification through a
  nanofluidic diode.
\newblock {\em Phys. Rev. E}, 76(4):041202, 2007.

\bibitem{vlassiouk_nl_2008}
I.~Vlassiouk, S.~Smirnov, and Z.~Siwy.
\newblock Ionic selectivity of single nanochannels.
\newblock {\em Nano Lett.}, 8(7):1978--1985, 2008.

\bibitem{song_plosone_2011}
C.~Song and B.~Corry.
\newblock Testing the applicability of {Nernst-Planck} theory in ion
  channels:comparisons with brownian dynamics simulations.
\newblock {\em {PLoS} {ONE}}, 6(6):e21204, 2011.

\bibitem{cervera_ea_2011}
J.~Cervera, P.~Ram{\'i}rez, S.~Mafe, and P.~Stroeve.
\newblock Asymmetric nanopore rectification for ion pumping, electrical power
  generation, and information processing applications.
\newblock {\em Electrochim. Acta}, 56(12):4504--4511, 2011.

\bibitem{siwy_prl_2002}
Z.~Siwy and A.~Fulinski.
\newblock Fabrication of a synthetic nanopore ion pump.
\newblock {\em Phys. Rev. Lett.}, 89(19):198103, 2002.

\bibitem{siwy_ss_2003}
Z.~Siwy, P.~Apel, D.~Baur, D.~D. Dobrev, Y.~E. Korchev, R.~Neumann, R.~Spohr,
  C.~Trautmann, and K.~O. Voss.
\newblock Preparation of synthetic nanopores with transport properties
  analogous to biological channels.
\newblock {\em Surf. Sci.}, 532:1061--1066, 2003.

\bibitem{siwy_nim_2003}
Z.~Siwy, P.~Apel, D.~Dobrev, R.~Neumann, R.~Spohr, C.~Trautmann, and K.~Voss.
\newblock Ion transport through asymmetric nanopores prepared by ion track
  etching.
\newblock {\em Nuclear Instruments \& Methods In Phys. Research Section B-beam
  Interactions With Materials Atoms}, 208:143--148, 2003.

\bibitem{howorka_siwy_chapt11_2009}
S.~Howorka and Z.~Siwy.
\newblock Nanopores: Generation, engineering, and singly-molecule applications.
\newblock In P.~Hinterdorfer and A.~van Oijen, editors, {\em Handbook of
  Single-Molecule Biophysics}, Advances in Chemical Physics, chapter Chapter
  11, pages 293--339. Springer, 2009.

\bibitem{guo_acr_2013}
W.~Guo, Y.~Tian, and L.~Jiang.
\newblock Asymmetric ion transport through ion-channel-mimetic solid-state
  nanopores.
\newblock {\em Acc. Chem. Res.}, 46(12):2834--2846, 2013.

\bibitem{gibb_chapter_2013}
T.~Gibb and M.~Ayub.
\newblock Solid-state nanopore fabrication.
\newblock In {\em Engineered Nanopores for Bioanalytical Applications}, pages
  121--140. Elsevier {BV}, 2013.

\bibitem{guan_nanotech_2014}
W.~Guan, S.~X. Li, and M.~A. Reed.
\newblock Voltage gated ion and molecule transport in engineered nanochannels:
  theory, fabrication and applications.
\newblock {\em Nanotechnology}, 25(12):122001, 2014.

\bibitem{zhang_nanotoday_2016}
Huacheng Zhang, Ye~Tian, and Lei Jiang.
\newblock Fundamental studies and practical applications of bio-inspired smart
  solid-state nanopores and nanochannels.
\newblock {\em Nano Today}, 8(3):1470--1478, 2016.

\bibitem{otto_chapter_2013}
O.~Otto and U.~F. Keyser.
\newblock {DNA} translocation.
\newblock In {\em Engineered Nanopores for Bioanalytical Applications}, pages
  31--58. Elsevier {BV}, 2013.

\bibitem{wang_cs_2017}
L.~Wang, Y.~Feng, Y.~Zhou, M.~Jia, G.~Wang, W.~Guo, and L.~Jiang.
\newblock Photo-switchable two-dimensional nanofluidic ionic diodes.
\newblock {\em Chem. Sci.}, 8(6):4381--4386, 2017.

\bibitem{sexton_mbs_2007}
L.~T. Sexton, L.~P. Horne, and C.~R. Martin.
\newblock Developing synthetic conical nanopores for biosensing applications.
\newblock {\em Mol. BioSyst.}, 3:667--685, 2007.

\bibitem{howorka_csr_2009}
S.~Howorka and Z.~Siwy.
\newblock Nanopore analytics: sensing of single molecules.
\newblock {\em Chem. Soc. Rev.}, 38(8):2360--2384, 2009.

\bibitem{piruska_csr_2010}
A.~Piruska, M.~Gong, and J.~V. Sweedler.
\newblock Nanofluidics in chemical analysis.
\newblock {\em Chem. Soc. Rev.}, 39:1060--1072, 2010.

\bibitem{howorka_nbt_2012}
S.~Howorka and Z.~S. Siwy.
\newblock Nanopores as protein sensors.
\newblock {\em Nat. Biotechnol.}, 30(6):506--507, jun 2012.

\bibitem{daiguji_nl_2005}
H.~Daiguji, Y.~Oka, , and K.~Shirono.
\newblock Nanofluidic diode and bipolar transistor.
\newblock {\em Nano Letters}, 5(11):2274--2280, 2005.

\bibitem{karnik_nl_2007}
R.~Karnik, C.~Duan, K.~Castelino, H.~Daiguji, and A.~Majumdar.
\newblock Rectification of ionic current in a nanofluidic diode.
\newblock {\em Nano Lett.}, 7(3):547--551, 2007.

\bibitem{vlassiouk_acsnanno_2008}
I.~Vlassiouk, S.~Smirnov, and Z.~Siwy.
\newblock Nanofluidic ionic diodes. comparison of analytical and numerical
  solutions.
\newblock {\em ACS Nano}, 2(8):1589--1602, 2008.

\bibitem{yan_nl_2009}
R.~Yan, W.~Liang, R.~Fan, and P.~Yang.
\newblock Nanofluidic diodes based on nanotube heterojunctions.
\newblock {\em Nano Lett.}, 9(11):3820--3825, 2009.

\bibitem{nguyen_nt_2010}
G.~Nguyen, I.~Vlassiouk, and Z.~S Siwy.
\newblock Comparison of bipolar and unipolar ionic diodes.
\newblock {\em Nanotech.}, 21(26):265301, 2010.

\bibitem{szymczyk_jpcb_2010}
A.~Szymczyk, H.~Zhu, and B.~Balannec.
\newblock Ion rejection properties of nanopores with bipolar fixed charge
  distributions.
\newblock {\em J. Phys. Chem. B}, 114(31):10143--10150, aug 2010.

\bibitem{singh_jap_2011}
K.~P. Singh and M.~Kumar.
\newblock Effect of surface charge density and electro-osmotic flow on ionic
  current in a bipolar nanopore fluidic diode.
\newblock {\em J. Appl. Phys.}, 110(8):084322, 2011.

\bibitem{singh_jpcb_2011}
K.~P. Singh and M.~Kumar.
\newblock Effect of nanochannel diameter and debye length on ion current
  rectification in a fluidic bipolar diode.
\newblock {\em J. Phys. Chem. C}, 115(46):22917--22924, 2011.

\bibitem{singh_apl_2011}
K.~P. Singh, K.~Kumari, and M.~Kumar.
\newblock Ion current rectification in a fluidic bipolar nanochannel with
  smooth junction.
\newblock {\em Appl. Phys. Lett.}, 99(11):113103, 2011.

\bibitem{van_oeffelen_plosone_2015}
L.~van Oeffelen, W.~Van~Roy, H.~Idrissi, D.~Charlier, L.~Lagae, and G.~Borghs.
\newblock Ion current rectification, limiting and overlimiting conductances in
  nanopores.
\newblock {\em {PLOS} {ONE}}, 10(5):e0124171, may 2015.

\bibitem{tajparast_bba_2015}
M.~Tajparast, G.~Virdi, and M.~I. Glavinovi\'{c}.
\newblock Spatial profiles of potential, ion concentration and flux in short
  unipolar and bipolar nanopores.
\newblock {\em Biochim. Biophys. Acta (BBA) - Biomem.}, 1848(10, Part
  A):2138--2153, 2015.

\bibitem{hato-cmp-19-13802-2016}
Z.~Hat\'o, D.~Boda, D.~Gillepie, J.~Vrabec, G.~Rutkai, and T.~Krist\'of.
\newblock Simulation study of a rectifying bipolar ion channel: detailed model
  versus reduced model.
\newblock {\em Cond. Matt. Phys.}, 19(1):13802, 2016.

\bibitem{matejczyk-jcp-submitted-2017}
B.~Matejczyk, M.~Valisk\'o, M.-T. Wolfram, J.-F. Pietschmann, and D.~Boda.
\newblock Multiscale modeling of a rectifying bipolar nanopore: {Comparing}
  {Poisson-Nernst-Planck} to {Monte Carlo}.
\newblock {\em J. Chem. Phys.}, 146(12):124125, 2017.

\bibitem{aksimentiev_ieee_2009}
A.~Aksimentiev, R.~K. Brunner, E.~Cruz-Ch\'u, J.~Comer, and K.~Schulten.
\newblock Modeling transport through synthetic nanopores.
\newblock {\em IEEE Nanotechnol. Mag.}, 3(1):20--28, 2009.

\bibitem{cruzchu_fd_2009}
E.~R. Cruz-Chu, T.~Ritz, Z.~S. Siwy, and K.~Schulten.
\newblock Molecular control of ionic conduction in polymer nanopores.
\newblock {\em Faraday Discussions}, 143:47--62, 2009.

\bibitem{cruzchu_jpcc_2009}
E.~R. Cruz-Chu, A.~Aksimentiev, and K.~Schulten.
\newblock Ionic current rectification through silica nanopores.
\newblock {\em J. Phys. Chem. C}, 113(5):1850--1862, 2009.

\bibitem{chen_small_2011}
Q.~Chen, L.~Meng, Q.~Li, D.~Wang, W.~Guo, Z.~Shuai, and L.~Jiang.
\newblock Water transport and purification in nanochannels controlled by
  asymmetric wettability.
\newblock {\em Small}, 7(15):2225--2231, 2011.

\bibitem{gamble_jpcc_2014}
T.~Gamble, K.~Decker, T.~S Plett, M.~Pevarnik, J.-F. Pietschmann, I.~V.
  Vlassiouk, A.~Aksimentiev, and Z.~S. Siwy.
\newblock Rectification of ion current in nanopores depends on the type of
  monovalent cations -- experiments and modeling.
\newblock {\em J. Phys. Chem. C}, 118(18):9809--9819, 2014.

\bibitem{ge_ms_2016}
Y.~Ge, J.~Zhu, M.~Kang, J.~Xian, Q.~An, and G.~Zhong.
\newblock Ion current rectification in a confined conical nanopore with high
  solution concentrations.
\newblock {\em Mol. Sim.}, 42(11):942--947, 2016.

\bibitem{vmd}
W.~Humphrey, A.~Dalke, and K.~Schulten.
\newblock {VMD} -- {V}isual {M}olecular {D}ynamics.
\newblock {\em J. Mol. Graphics}, 14:33--38, 1996.

\bibitem{Berendsen199543}
H.~J.~C. Berendsen, D.~van~der Spoel, and R.~van Drunen.
\newblock {GROMACS}: {A} message-passing parallel molecular dynamics
  implementation.
\newblock {\em Comp. Phys. Comm.}, 91(1--3):43--56, 1995.

\bibitem{Pronk01042013}
S.~Pronk, Sz. P\'{a}ll, R.~Schulz, P.~Larsson, P.~Bjelkmar, R.~Apostolov, M.~R.
  Shirts, J.~C. Smith, P.~M. Kasson, D.~van~der Spoel, B.~Hess, and E.~Lindahl.
\newblock {GROMACS} 4.5: a high-throughput and highly parallel open source
  molecular simulation toolkit.
\newblock {\em Bioinformatics}, 29(7):845--854, 2013.

\bibitem{charmm}
P.~Bjelkmar, P.~Larsson, M.~A. Cuendet, B.~Hess, and E.~Lindahl.
\newblock Implementation of the {CHARMM} force field in {GROMACS}: {Analysis}
  of protein stability effects from correction maps, virtual interaction sites,
  and water models.
\newblock {\em J. Chem. Theor. Comp.}, 6(2):459--466, 2010.

\bibitem{bussi_jcp_2007}
G.~Bussi, D.~Donadio, and M.~Parrinello.
\newblock Canonical sampling through velocity rescaling.
\newblock {\em J. Chem. Phys.}, 126(1):014101, 2007.

\end{thebibliography}

\end{document}